\documentclass[final,5p,times,twocolumn,nopreprintline]{elsarticle}
\usepackage[T1]{fontenc}
\usepackage{amsmath,amssymb,microtype,graphicx,booktabs}
\usepackage{tikz,pgfplots}
\usetikzlibrary{decorations.pathmorphing,decorations.markings,arrows.meta}
\pgfplotsset{compat=1.18}
\usepackage[colorlinks=true,linkcolor=blue,citecolor=blue,urlcolor=blue]{hyperref}
\biboptions{numbers,sort&compress}
\journal{Physics Letters B}
\hypersetup{pdftitle={Flavour Alignment and Operator Sensitivity in a Left--Right Universal-Seesaw Model},pdfauthor={Claudio Coriano and Paul H. Frampton}}
\begin{document}
\begin{frontmatter}
\title{\textbf{Flavor Alignment and Operator Sensitivity\\in a Left--Right Universal-Seesaw Model}}
\author[a,b,c]{Claudio Corian\`o}
\author[a]{Paul H. Frampton}
\affiliation[a]{organization={Dipartimento di Matematica e Fisica, Universit\`a del Salento},
 addressline={Via Arnesano},city={Lecce},postcode={73100},country={Italy}}
\affiliation[b]{organization={INFN Sezione di Lecce},
 addressline={Via Arnesano},city={Lecce},postcode={73100},country={Italy}}
\affiliation[c]{organization={CNR-Nanotec},city={Lecce},country={Italy}}
\begin{abstract}
We analyse the flavour structure and low-energy operator sensitivity of a parity-related doublet universal-seesaw model with an imposed lepton triality. Degenerate bare singlet masses permit a basis in which light-light tree neutral flavour transitions vanish; the physical heavy masses remain nondegenerate. We connect this alignment property to the axial and scalar differences measured by $B_{d,s}^0\to\mu^+\mu^-$ and reproduce the six-state charged-vector mixing amplitudes for five reference masses. The reference analysis identifies kaon CP violation as the more sensitive flavour probe. Historical kaon data disfavour the lower reference masses, while estimates suggest that dimuon and radiative $B$-decay effects are harder to resolve at present precision. These comparisons use the specified reference spectrum and historical inputs. We also compare charged and neutral resonance rates with published search benchmarks. The vacuum hierarchy entails a cancellation of about $5\times10^{-6}$ at $M_{W_R}=80\,\mathrm{TeV}$. The initially thermal Dirac-neutrino branch has a dataset-dependent extra-radiation tension, strengthened in an illustrative mass-dependent decoupling estimate. For dimuon decays we establish operator selection rules and sensitivity estimates. The numerical flavour results cover meson mixing and charged-vector dipoles. We give no dimuon branching-fraction prediction or global allowed region.
\end{abstract}
\end{frontmatter}
\section{Introduction}
Neutral flavour transitions distinguish the flavour structure of a left--right theory from its gauge-group assignment. Tree-level neutral-vector exchange requires a flavour-changing vertex generated by the fermion rotations. The pseudoscalar meson then selects the axial and scalar combinations of the semileptonic coefficients. These facts motivate an operator and flavour-structure analysis of the doublet universal-seesaw model.

We adopt the established model and flavour kernels of Ref.~\cite{Dcruz}. The parity-based tree strong-CP mechanism and radiative mass-phase analysis are established results~\cite{Hisano,Banno}, while the entropy relation for thermal right-handed neutrinos is taken from Ref.~\cite{Luo}. These constructions, kernels and the radiation-floor relation are established inputs drawn from the cited literature. Our contribution is to organise the degenerate-singlet limit through exact mass rotations and dimuon selection rules, exhibit the resulting sensitivity hierarchy, and make the numerical reference trajectory and its consistency conditions explicit. The meson-mixing and dipole entries reproduce the existing reference calculation using those kernels. Reference~\cite{Colangelo2025} studies an Abelian gauge extension with different field content. Its correlated-coupling analysis provides context for the operator discussion. The LR dimuon loop matching remains to be calculated.

We use the dimuon data to quantify amplitude sensitivity and derive the operator combinations entering a future gauge-independent matching calculation. The numerical analysis gives complete reference tables, a signed historical diagnostic, and quantitative checks of collider rates, the scalar hierarchy and the thermal branch. The assumptions about running, decoupling and experimental inputs are stated with each calculation.

\section{Dimuon measurements and their interpretation}
The LHCb measurement of Ref.~\cite{LHCb} uses $18\,\mathrm{fb}^{-1}$ of Run~3 data and finds $3.0\sigma$ evidence for $B^0\to\mu^+\mu^-$. The combination of Runs~1--3 gives $3.2\sigma$ and
\begin{align}
 \mathcal B(B^0\to\mu^+\mu^-)
 &=\left(1.14^{+0.42+0.06}_{-0.39-0.05}\right)\times10^{-10},\\
 \overline{\mathcal B}(B_s^0\to\mu^+\mu^-)
 &=\left(3.25^{+0.26+0.13}_{-0.25-0.12}\right)\times10^{-9},\\
 \tau_{\mu\mu}&=1.60\pm0.16\pm0.01\,\mathrm{ps}.
\end{align}
The uncertainties are statistical and systematic. For the comparison we use the SM benchmarks quoted in Ref.~\cite{LHCb},
\begin{align}
 \overline{\mathcal B}_{s,\rm SM}&=(3.66\pm0.14)\times10^{-9},\\
 \mathcal B_{d,\rm SM}&=(1.03\pm0.05)\times10^{-10}.
 \label{eq:SMdimuon}
\end{align}
The measured $B_s$ central value lies below this reference: adding its upward experimental errors and the SM error in quadrature gives a pull of about $-1.27$. The $B_d$ central value is above its reference by about $0.28$ standard deviations using the downward experimental errors. The quoted pulls use one-dimensional Gaussian error propagation; a joint likelihood also requires the experimental correlations. Agreement with the SM is therefore maintained. The $B_s$ benchmark is the time-integrated one used below. The upper systematic error $+0.06$ in the combined $B_d$ result follows the displayed combination in Ref.~\cite{LHCb}, v2; it is retained after checking that version. The reported signal significance measures the evidence for the decay. Testing its agreement with the SM requires the measured rate and its uncertainties. We write a bar on the $B_s$ branching fraction to emphasise the experimental time integration.

CMS has independently measured the $B_s^0\to\mu^+\mu^-$ branching fraction and effective lifetime and searched for the $B^0$ mode~\cite{CMSdimuon}. Its published 2016--2018 analysis found agreement with the SM and no evidence for the latter decay. The experimental likelihoods can enter a joint flavour analysis when their correlations and normalisation inputs are treated consistently.

\section{A four-dimensional left--right framework for flavour}
We adopt the conventional left--right gauge structure~\cite{MP,SM}, with QCD colour included explicitly,
\begin{equation}
 G_{\rm LR}=SU(3)_c\times SU(2)_L\times SU(2)_R\times U(1)_{B-L}.
\end{equation}
For each generation the ordinary fermions transform as
\begin{align}
 Q_L&\sim(3,2,1,\tfrac13),& Q_R&\sim(3,1,2,\tfrac13),\\
 L_L&\sim(1,2,1,-1),& L_R&\sim(1,1,2,-1),
\end{align}
The representation labels list the dimensions under $SU(3)_c$, $SU(2)_L$, $SU(2)_R$, followed by the charge $B-L$. Here $L_R=(\nu_R,e_R)^T$. Electric charge is $Q_{\rm em}=T_L^3+T_R^3+t_{BL}$, with $t_{BL}=(B-L)/2$. The labels $L$ and $R$ denote chirality and weak representations in the same spacetime.

Generalised parity exchanges the sectors together with ordinary spatial inversion, $Q_L(t,\mathbf x)\leftrightarrow\gamma^0Q_R(t,-\mathbf x)$, and similarly for leptons. It exchanges $W_L$ and $W_R$ with the vector transformation signs. The common Abelian field $X_\mu$ couples as
\begin{equation}
 \mathcal L_X=g_BX_\mu\sum_{f,\chi=L,R}
 t_{BL,f}\bar f\gamma^\mu P_\chi f,
 \qquad P_{L,R}=\frac{1\mp\gamma_5}{2}.
\end{equation}
We call this shared interaction a gauge portal between the chiral sectors. It introduces no new charge for an ordinary fermion. Its gauge couplings obey the parity relations at the symmetry scale; running and threshold effects must be treated when comparing low-energy parameters.

The scalar multiplets and neutrino-mass interactions complete the specification of the gauge theory. The conventional bidoublet and triplet realisation can generate heavy Majorana neutrinos. A doublet realisation with $\chi_L\sim(1,2,1,+1)$ and $\chi_R\sim(1,1,2,+1)$ admits universal-seesaw mass generation through additional vector-like fermions~\cite{Dcruz}. We use the doublet breaking pattern throughout the reference calculation. Writing $\langle\chi_{L,R}^0\rangle=v_{L,R}/\sqrt2$ and taking $v_R\gg v_L$ gives, at leading order,
\begin{equation}
 M_{W_R}=\frac{g_Rv_R}{2},\qquad
 M_{Z_R}=\frac{\sqrt{g_R^2+g_B^2}\,v_R}{2},\qquad
 \frac1{g_Y^2}=\frac1{g_R^2}+\frac1{g_B^2}.
 \label{eq:masses}
\end{equation}
The symmetry-breaking scale therefore correlates the two heavy vector masses. The charged-vector mass matrix is diagonal at tree level in this doublet realisation.

At the same order, setting $g_{RB}=\sqrt{g_R^2+g_B^2}$, the neutral fields are
\begin{equation}
 B_Y=\frac{g_BW_R^3+g_RX}{g_{RB}},\qquad
 Z_R=\frac{g_RW_R^3-g_BX}{g_{RB}}.
\end{equation}
The heavy state couples to $(g_R^2T_R^3-g_B^2t_{BL})/g_{RB}$. In particular,
\begin{equation}
 \begin{aligned}
 g_{\mu L}^{Z_R}&=\frac{g_B^2}{2g_{RB}},&
 g_{\mu R}^{Z_R}&=\frac{g_B^2-g_R^2}{2g_{RB}},\\
 g_{\mu A}^{Z_R}&\equiv\frac{g_{\mu R}^{Z_R}-g_{\mu L}^{Z_R}}2
 =-\frac{g_R^2}{4g_{RB}}.
 \end{aligned}
 \label{eq:axial}
\end{equation}
Even though the $B-L$ gauge eigenstate is vectorial for charged leptons, its physical heavy neutral combination is generally axial as well as vectorial.

Flavour generation is a separate requirement. With only the ordinary quarks, a universal neutral coupling remains diagonal under unitary mass rotations. The $B-L$ portal alone consequently supplies no tree-level $\bar q bZ_R$ vertex. Charged-current loops, scalar interactions or suitable mixing with additional fermions must provide the flavour transition. For example, a four-dimensional universal-seesaw mass matrix has the form
\begin{equation}
 \mathcal M_d=\begin{pmatrix}
 0&Y_{dL}v_L/\sqrt2\\Y_{dR}^\dagger v_R/\sqrt2&M_D
 \end{pmatrix},\qquad
 m_d\simeq-\frac{v_Lv_R}{2}Y_{dL}M_D^{-1}Y_{dR}^\dagger.
\end{equation}
Mixing states with different neutral generators can induce flavour-changing light-quark couplings. The same rotations modify charged currents and must reproduce quark masses and CKM data. A universal-seesaw completion is therefore an extension of the ordinary field content, with consequences beyond the desired beauty amplitude. Existing studies find restrictive correlations~\cite{Dcruz}.

Parity also motivates an approach to strong CP. We use the convention $\bar\theta=\theta_{\rm QCD}+\arg\det(\mathcal M_u\mathcal M_d)$. A consistent parity solution must control the bare angle, vacuum-induced mass phases and radiative effects. The neutron electric dipole moment remains a correlated test~\cite{Hisano}. The heavy-vector scale and neutral-lepton spectrum enter as additional model inputs.

\section{A specified scalar and Dirac-neutrino realisation}
We use the doublet branch with an explicit scalar potential and Dirac-neutrino mass sector. In addition to the ordinary doublets, introduce three vector-like copies of $U\sim(3,1,1,4/3)$, $D\sim(3,1,1,-2/3)$, $E\sim(1,1,1,-2)$ and $N\sim(1,1,1,0)$. The representation labels follow the ordering used above. At the level of the effective renormalisable theory, impose a lepton triality under which all ordinary leptons, $E$ and $N$ have charge one modulo three, while quarks and scalars are neutral. The Yukawa interactions and Dirac singlet masses preserve this symmetry, whereas Majorana bilinears carry charge two and are forbidden. Lepton triality supplements the LR gauge symmetry as an assumption of the reference theory. Removing Majorana terms leaves a classical continuous lepton number, but its electroweak anomaly changes lepton number by three units for three families. Its $Z_3$ subgroup survives those instantons and is enough to forbid the Majorana bilinears. In left-handed Weyl notation the mixed $[SU(2)_L]^2Z_3$ and $[SU(2)_R]^2Z_3$ anomaly coefficients, with $2T(\mathbf2)=1$, are $+3$ and $-3$, hence zero modulo three. The colour coefficient vanishes; the gravitational and cubic sums, and the mixed Abelian sums involving $B-L$, cancel between opposite-chirality partners. The extra singlets are vector-like in these sums. The listed anomaly sums are consistent with lepton triality in this field content. Embedding it in a discrete-gauge ultraviolet theory and controlling higher-dimensional lepton-number violation require additional assumptions. We impose triality at the level of the effective renormalisable theory.

At a common reference scale the parity-related Yukawa interactions have the form
\begin{align}
 -\mathcal L_Y={}&\bar Q_LY_u\widetilde\chi_LU_R+
 \bar Q_RY_u\widetilde\chi_RU_L+
 \bar Q_LY_d\chi_LD_R+\bar Q_RY_d\chi_RD_L\nonumber\\
 &+\bar L_LY_e\chi_LE_R+\bar L_RY_e\chi_RE_L+
 \bar L_LY_\nu\widetilde\chi_LN_R+
 \bar L_RY_\nu\widetilde\chi_RN_L\nonumber\\
 &+
 \sum_{F=U,D,E,N}\bar F_LM_FF_R+\mathrm{h.c.},
\end{align}
with Hermitian $M_F$ and $\widetilde\chi=i\sigma_2\chi^*$. Parity breaking is confined at tree level to real dimension-two scalar parameters. We take
\begin{equation}
 V=-\mu_L^2\chi_L^\dagger\chi_L-\mu_R^2\chi_R^\dagger\chi_R
 +\lambda\big[(\chi_L^\dagger\chi_L)^2+(\chi_R^\dagger\chi_R)^2\big]
 +\kappa(\chi_L^\dagger\chi_L)(\chi_R^\dagger\chi_R).
 \label{eq:potential}
\end{equation}
The soft term $\mu_L^2-\mu_R^2$ permits $v_R\gg v_L>0$. Stationarity requires $\mu_L^2=\lambda v_L^2+\kappa v_R^2/2$ and $\mu_R^2=\lambda v_R^2+\kappa v_L^2/2$. The radial scalar mass matrix is
\begin{equation}
 \mathcal M_h^2=\begin{pmatrix}2\lambda v_L^2&\kappa v_Lv_R\\
 \kappa v_Lv_R&2\lambda v_R^2\end{pmatrix}.
\end{equation}
For $\lambda>0$ and $2\lambda>|\kappa|$, the stationary point with both expectation values nonzero is the global tree minimum. Indeed, with $x_A=\chi_A^\dagger\chi_A$ and $\delta x_A=x_A-v_A^2/2$, stationarity gives
\begin{equation}
 V-V_{\min}=\lambda\delta x_L^2+\lambda\delta x_R^2
 +\kappa\delta x_L\delta x_R>0
 \label{eq:globalvacuum}
\end{equation}
away from the minimum in $(x_L,x_R)$. The quadratic form has eigenvalues $\lambda\pm\kappa/2>0$ on the whole real plane, hence also on $x_L,x_R\geq0$. Gauge-equivalent orientations represent the same vacuum. The two neutral vacuum phases can be removed by gauge transformations; there is no physical vacuum CP phase in this doublet-only scalar sector. These vacuum conditions hold at tree level. Their scale dependence requires a running stability analysis. The hierarchy has a cost in this scalar parametrisation. Define the cancellation fraction
\begin{equation}
 \epsilon_{\rm vac}=\frac{\lambda v_L^2}{\mu_L^2}
 =\frac{\lambda v_L^2}{\lambda v_L^2+\kappa v_R^2/2}.
 \label{eq:vacuumtuning}
\end{equation}
It is $1.38\times10^{-3}$ at $5\,\mathrm{TeV}$ and $5.41\times10^{-6}$ at $80\,\mathrm{TeV}$ for the stated scan. Equivalently, $\mu_L^2$ and $\kappa v_R^2/2$ must cancel in the light-scale stationarity equation at this relative accuracy. Soft parity breaking permits the hierarchy with this explicit cancellation. The naturalness of the hierarchy remains unexplained by the soft parity-breaking term. The quantity $\epsilon_{\rm vac}$ measures the cancellation in the chosen scalar parametrisation.

For each fermion sector write $U_{fL}^\dagger\mathcal M_fU_{fR}=\widehat m_f$, where
\begin{equation}
 \mathcal M_f=\begin{pmatrix}0&Y_fv_L/\sqrt2\\Y_f^\dagger v_R/\sqrt2&M_F\end{pmatrix},
 \qquad f=u,d,e,\nu.
 \label{eq:commonfermion}
\end{equation}
We use the aligned reference texture
\begin{equation}
 \begin{aligned}
 Y_u&=\widehat y_u,& Y_d&=V\widehat y_d,\\
 Y_e&=\widehat y_e,& Y_\nu&=U\widehat y_\nu,\\
 M_F&=M\mathbf1.
 \end{aligned}
 \label{eq:alignment}
\end{equation}
Here $V$ carries the CKM phase and $U$ is an illustrative PMNS matrix. At a fixed matching scale, the singular-value decomposition $Y_f=L_f\widehat y_fR_f^\dagger$ allows $R_f$ to be absorbed into identical rotations of $F_L,F_R$ when $M_F\propto\mathbf1$. Relative ordinary-quark rotations remain in $V$. The aligned Yukawa form therefore expresses the basis-independent FCNC cancellation implied by degenerate singlet masses and the stated parity relations. A common value $M$ across different charge sectors is an additional simplification. Generic nondegenerate singlet masses or a change of boundary scale can destroy the alignment. In our Lagrangian convention parity equates the two Yukawa vertices; the lower-left mass block contains $Y_f^\dagger$, as in Eq.~\eqref{eq:commonfermion}. After a unitary flavour rotation each mass matrix reduces to three real $2\times2$ blocks $(\begin{smallmatrix}0&a_i\\b_i&M\end{smallmatrix})$, with $a_i=y_iv_L/\sqrt2$ and $b_i=y_iv_R/\sqrt2$. Given a desired light singular value $m_i$, the Yukawa coupling is obtained exactly from
\begin{equation}
 p y_i^4-m_i^2 k y_i^2+m_i^4-m_i^2M^2=0,
 \qquad p=\frac{v_L^2v_R^2}{4},\qquad k=\frac{v_L^2+v_R^2}{2}.
 \label{eq:yexact}
\end{equation}
The heavy singular value is $\sqrt{M^2+k y_i^2-m_i^2}$ for the positive solution. We retain the exact mixing angles, including the top sector, for which an expansion in $Y_uv_R/M$ is inappropriate.

Let $\Pi=\operatorname{diag}(1,1,1,0,0,0)$ project onto ordinary doublets. The complete charged matrices are $\mathbb V_A=U_{uA}^\dagger\Pi U_{dA}$ and $\mathbb K_A=U_{eA}^\dagger\Pi U_{\nu A}$. The projection onto the light states determines the departures of their light blocks from unitarity. For any neutral gauge generator $T_A$, its physical coupling is $U_{fA}^\dagger T_AU_{fA}$; scalar couplings follow by rotating $\partial\mathcal M_f/\partial v_L$ and $\partial\mathcal M_f/\partial v_R$. Thus neutral couplings, charged couplings and scalar vertices come from the same rotations. In the texture~\eqref{eq:alignment}, light-light neutral and scalar couplings are diagonal within each charge sector, even though $V$ is complex. Consequently the new tree-level coefficients in Eq.~\eqref{eq:neutralmatching} and the corresponding tree scalar coefficients vanish. Charged-current loop contributions survive.

In the neutral basis $(W_L^3,W_R^3,X)$ we diagonalise
\begin{equation}
 \mathcal M_0^2=\frac14\begin{pmatrix}
 g_L^2v_L^2&0&-g_Lg_Bv_L^2\\
 0&g_R^2v_R^2&-g_Rg_Bv_R^2\\
 -g_Lg_Bv_L^2&-g_Rg_Bv_R^2&g_B^2(v_L^2+v_R^2)
 \end{pmatrix}.
 \label{eq:exactneutral}
\end{equation}
The photon is massless and the other eigenstates are denoted $Z_1,Z_2$. Both poles are retained when matching low-energy neutral operators. The tree charged matrix is diagonal. Radiative $W_L$--$W_R$ self-energies supply a mixing correction beyond this tree-level mass matrix.

\section{Beauty amplitudes and neutral-current matching}
The $B^0$ measurement concerns the $b\to d$ sector, whereas $B_s^0$ probes $b\to s$. Both must be retained. For $q=d,s$, define
\begin{equation}
 \begin{aligned}
 \mathcal H_{b\to q\mu\mu}={}&-\mathcal N_q\sum_{i=9,10,S,P}
 \big(C_i^{(q)}Q_i^{(q)}+C_i^{\prime(q)}Q_i^{\prime(q)}\big)\\
 &+\cdots+\mathrm{h.c.},\\
 \mathcal N_q={}&\frac{G_F\alpha_e}{\sqrt2\pi}V_{tb}V_{tq}^*.
 \end{aligned}
\end{equation}
The operators are
\begin{align}
 Q_9^{(q)}&=(\bar q\gamma_\alpha P_Lb)(\bar\mu\gamma^\alpha\mu),&
 Q_{10}^{(q)}&=(\bar q\gamma_\alpha P_Lb)(\bar\mu\gamma^\alpha\gamma_5\mu),\\
 Q_S^{(q)}&=(\bar qP_Rb)(\bar\mu\mu),&
 Q_P^{(q)}&=(\bar qP_Rb)(\bar\mu\gamma_5\mu).
\end{align}
Primed operators interchange the quark projectors. No quark-mass factor is included in $Q_{S,P}$. The ellipsis includes the dipole, four-quark and other terms needed in a complete calculation. Coefficients and masses must be evaluated in compatible schemes and at specified scales.

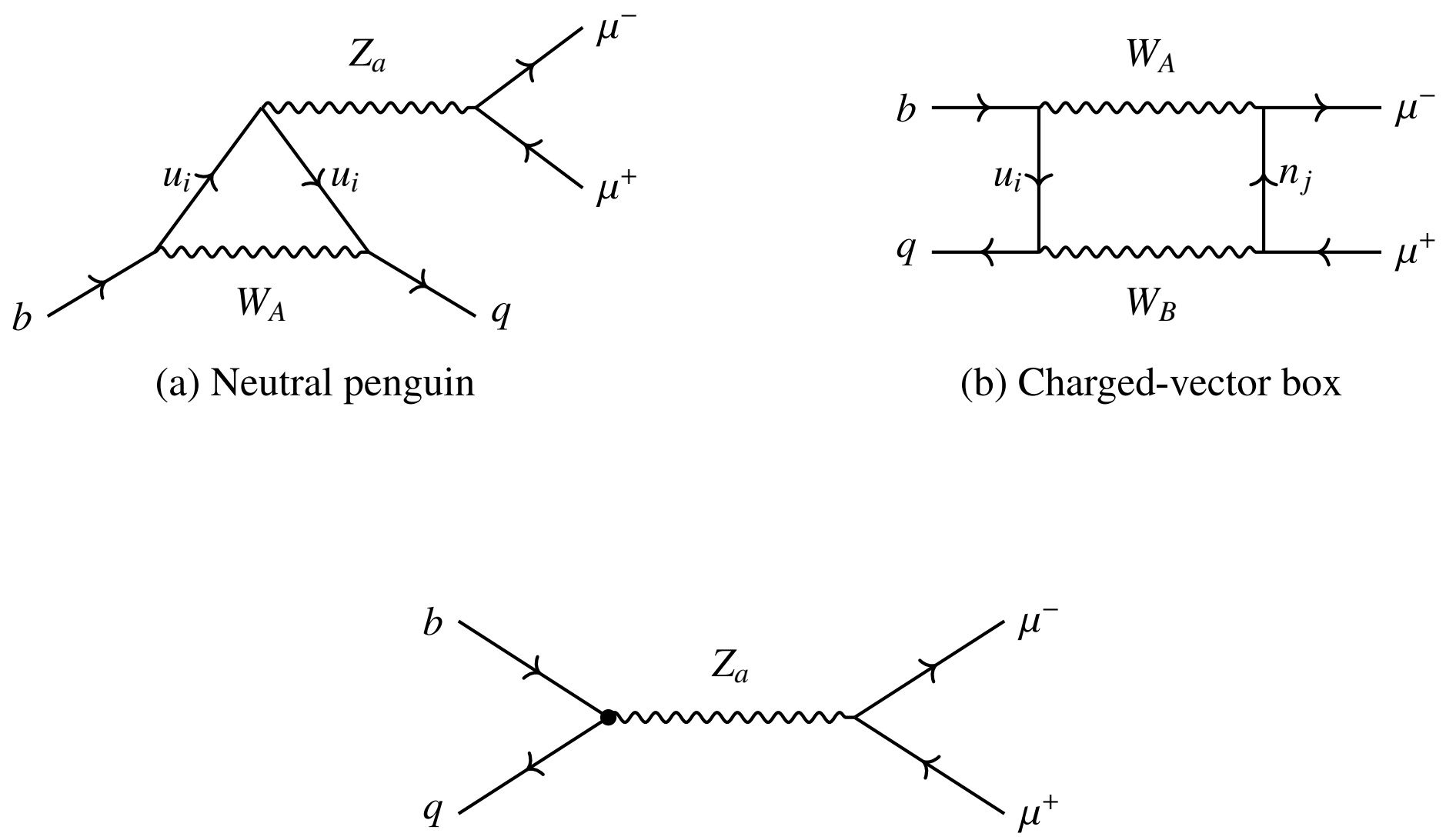
\begin{figure*}[t]
\centering
\begin{tikzpicture}[x=.92cm,y=.92cm,
 fermion/.style={thick,postaction={decorate},decoration={markings,mark=at position .55 with {\arrow{>}}}},
 vector/.style={thick,decorate,decoration={snake,amplitude=1.2pt,segment length=5pt}}]
% Neutral penguin: quark-level b -> q mu+ mu-.
\begin{scope}
 \draw[fermion] (-1,-.6) node[left] {$b$} -- (0,0);
 \draw[fermion] (0,0) -- (1,1.35) node[midway,left] {$u_i$};
 \draw[fermion] (1,1.35) -- (2,0) node[midway,right] {$u_i$};
 \draw[fermion] (2,0) -- (3,-.6) node[right] {$q$};
 \draw[vector] (0,0) -- (2,0) node[midway,below=5pt] {$W_A$};
 \draw[vector] (1,1.35) -- (3,1.35) node[midway,above=5pt] {$Z_a$};
 \draw[fermion] (3,1.35) -- (4,2.1) node[right] {$\mu^-$};
 \draw[fermion] (4,.6) node[right] {$\mu^+$} -- (3,1.35);
 \node at (1.5,-1.25) {(a) Neutral penguin};
\end{scope}
% Charged box, crossed to the annihilation channel.
\begin{scope}[xshift=7.6cm]
 \draw[fermion] (-1,1.35) node[left] {$b$} -- (0,1.35);
 \draw[fermion] (0,0) -- (-1,0) node[left] {$q$};
 \draw[fermion] (0,1.35) -- (0,0) node[midway,left] {$u_i$};
 \draw[vector] (0,1.35) -- (2.1,1.35) node[midway,above=5pt] {$W_A$};
 \draw[vector] (0,0) -- (2.1,0) node[midway,below=5pt] {$W_B$};
 \draw[fermion] (2.1,0) -- (2.1,1.35) node[midway,right] {$n_j$};
 \draw[fermion] (2.1,1.35) -- (3.2,1.35) node[right] {$\mu^-$};
 \draw[fermion] (3.2,0) node[right] {$\mu^+$} -- (2.1,0);
 \node at (1.05,-1.25) {(b) Charged-vector box};
\end{scope}
% Possible tree FCNC: explicitly absent in aligned benchmark.
\begin{scope}[xshift=3.9cm,yshift=-4.0cm]
 \draw[fermion] (-1.4,.9) node[left] {$b$} -- (0,0);
 \draw[fermion] (0,0) -- (-1.4,-.9) node[left] {$q$};
 \draw[vector] (0,0) -- (2.3,0);
 \node at (1.15,.48) {$Z_a$};
 \draw[fermion] (2.3,0) -- (3.7,.9) node[right] {$\mu^-$};
 \draw[fermion] (3.7,-.9) node[right] {$\mu^+$} -- (2.3,0);
 \fill (0,0) circle (2pt);
 \node at (1.15,-1.5) {(c) Tree FCNC: zero in the aligned texture};
\end{scope}
\end{tikzpicture}
\caption{Representative short-distance topologies for $b\to q\mu^+\mu^-$, $q=d,s$, crossed as needed to obtain $\bar B_q^0=(b\bar q)\to\mu^+\mu^-$. Here $A,B=L,R$, $Z_a=Z_1,Z_2$, and the internal indices run over all six physical states of the corresponding fermion sector. Charge-conjugate diagrams describe $B_q^0$. Panel (a) shows neutral emission from the internal quark; emission from the charged line and the gauge-completing Goldstone and external-leg terms are also required. The chiral projectors select the contributing assignments in panel (b). Panel (c) illustrates neutral-vector exchange with an effective flavour vertex; it vanishes at tree level in the chosen aligned spectrum, as does the corresponding neutral-scalar exchange. These topologies specify the dimuon matching calculation. The numerical scan below evaluates charged-vector meson-mixing boxes and radiative dipoles.}
\label{fig:beautygraphs}
\end{figure*}

The diagrams in Fig.~\ref{fig:beautygraphs} organise the short-distance calculation in the specified LR theory. Charged-vector penguins and boxes involve the complete up-quark and neutral-lepton spectra, including the vector-like partners. Their separation into individual graphs is gauge dependent; Goldstone, external-leg and counterterm contributions must accompany them in a gauge-independent matching calculation. The scalar sector must be treated with the same fermion rotations. The neutral-exchange panel illustrates the flavour vertex of a general completion; its tree value vanishes in the aligned reference texture.

For physical neutral mass eigenstates $V_a$ with interactions
\begin{equation}
 \begin{aligned}
 \mathcal L_{\rm int}\supset\sum_a V_{a\alpha}\big[&g_{qb,L}^a\bar q\gamma^\alpha P_Lb\\
 &+g_{qb,R}^a\bar q\gamma^\alpha P_Rb+\mathrm{h.c.}\\
 &+g_{\mu L}^a\bar\mu\gamma^\alpha P_L\mu\\
 &+g_{\mu R}^a\bar\mu\gamma^\alpha P_R\mu\big].
 \end{aligned}
\end{equation}
tree exchange gives
\begin{equation}
 \begin{pmatrix}\Delta C_9^{(q)}\\\Delta C_{10}^{(q)}\end{pmatrix}
 =-\frac1{2\mathcal N_q}\sum_a\frac{g_{qb,L}^a}{M_a^2}
 \begin{pmatrix}g_{\mu L}^a+g_{\mu R}^a\\g_{\mu R}^a-g_{\mu L}^a\end{pmatrix},
 \label{eq:neutralmatching}
\end{equation}
with primed shifts obtained by replacing $g_{qb,L}^a$ by $g_{qb,R}^a$. The fermion rotations of the chosen completion determine all the couplings entering Eq.~\eqref{eq:neutralmatching}. Mixing-induced couplings of the light $Z$ must also be included. For thresholds above electroweak breaking, gauge-invariant matching and subsequent evolution are required.

Purely leptonic decays select a different combination from semileptonic angular observables. In the scalar normalisation above, introduce
\begin{align}
 P_q&=\frac{C_{10}^{(q)}-C_{10}^{\prime(q)}}{C_{10,\rm SM}^{(q)}}
 +\frac{M_{B_q}^2}{2m_\mu(m_b+m_q)}
 \frac{C_P^{(q)}-C_P^{\prime(q)}}{C_{10,\rm SM}^{(q)}},\\
 S_q&=\beta_q\frac{M_{B_q}^2}{2m_\mu(m_b+m_q)}
 \frac{C_S^{(q)}-C_S^{\prime(q)}}{C_{10,\rm SM}^{(q)}},\qquad
 \beta_q=\sqrt{1-\frac{4m_\mu^2}{M_{B_q}^2}}.
\end{align}
Within these operators the rate at production, normalised to its SM value, is $|P_q|^2+|S_q|^2$, with tiny SM scalar terms neglected. The vector muon current vanishes in the two-body projection at this order. The dimuon rate consequently selects the axial and scalar coefficients, while $C_9$ enters the semileptonic observables. The complementary sensitivity of semileptonic angular observables and dimuon rates to unprimed and primed Wilson coefficients has also been examined in a custodially protected Randall--Sundrum model~\cite{Biancofiore2014}.

The tree neutral-current selection rule can be written directly in terms of physical vertices:
\begin{equation}
 \Delta(C_{10}^{(q)}-C_{10}^{\prime(q)})
 =-\frac1{2\mathcal N_q}\sum_a
 \frac{(g_{qb,L}^a-g_{qb,R}^a)(g_{\mu R}^a-g_{\mu L}^a)}{M_a^2}.
 \label{eq:axialselection}
\end{equation}
A purely vectorial muon coupling sets the $Q_{10}$ contribution to zero. Equal left- and right-handed flavour vertices also cancel in the pseudoscalar-meson annihilation amplitude, including for an axial muon coupling. The flavour texture and symmetry breaking determine these coupling relations. In our aligned reference the stronger statement $g_{qb,L}^a=g_{qb,R}^a=0$ holds at tree level. The charged-loop amplitude must therefore be matched before assigning a dimuon branching fraction to this reference spectrum.

Writing $P_q=1+\delta P_q$ displays the interference with the Standard Model:
\begin{equation}
 R_q^{\rm prod}-1=2\operatorname{Re}\delta P_q+|\delta P_q|^2+|S_q|^2.
 \label{eq:interferencecontrast}
\end{equation}
An amplitude-level prediction is therefore required to determine both the linear shift and its phase. The interference term determines how a new coefficient changes the total decay probability.

For the $B_s$ system the comparison must include time integration~\cite{DeBruyn}. Defining $y_s=\Delta\Gamma_s/(2\Gamma_s)$ and $\mathcal A_s=\mathcal A_{\Delta\Gamma_s}^{\mu\mu}$ gives
\begin{equation}
 \begin{aligned}
 \frac{\overline{\mathcal B}_s}{\overline{\mathcal B}_{s,\rm SM}}
 &=\frac{1+\mathcal A_s y_s}{1+y_s}(|P_s|^2+|S_s|^2),\\
 \tau_{\mu\mu}&=\frac{\tau_{B_s}}{1-y_s^2}
 \frac{1+2\mathcal A_s y_s+y_s^2}{1+\mathcal A_s y_s}.
 \end{aligned}
\end{equation}
where the reference SM value is $\mathcal A_s=+1$. The amplitude phases and the neutral-meson mixing phase determine $\mathcal A_s$. The effective lifetime therefore adds phase information to the branching-fraction measurement. Experimental lifetime-dependent efficiencies should be treated consistently when fitting new amplitudes.

The same flavour couplings generate
\begin{equation}
 \begin{aligned}
 \mathcal H_{\Delta B=2}={}&\sum_a\frac1{2M_a^2}
 \big[(g_{qb,L}^a)^2O_{LL}+(g_{qb,R}^a)^2O_{RR}\\
 &\qquad+2g_{qb,L}^ag_{qb,R}^aO_{LR}\big]+\mathrm{h.c.}
 \end{aligned}
\end{equation}
where $O_{AB}=(\bar q\gamma_\alpha P_A b)
(\bar q\gamma^\alpha P_B b)$ with colour contracted within each current. Complex squares, operator mixing and hadronic matrix elements must be retained. Charged-vector boxes and the scalar contributions of the completion also enter. Meson-mixing studies constrain the same flavour couplings that enter the decay coefficients~\cite{BMN}.

\section{Decoupling and present dimuon sensitivity}
\label{sec:sensitivity}
A useful sensitivity test is possible before the full matching is known. Along the fixed-coupling trajectory $M_F=v_R$ all new heavy masses grow with $v_R$, whereas the dimensionless Yukawa singular values tend to constants. Once the low-energy inputs are fixed, dimension-six effects decouple as $v_L^2/v_R^2$, up to logarithms and flavour factors. With $g_R=g_L$, define
\begin{equation}
 \epsilon_R^{(6)}=\frac{v_L^2}{v_R^2}=\frac{M_W^2}{M_{W_R}^2},\qquad
 \delta P_q=\epsilon_R^{(6)}\mathcal F_q.
 \label{eq:reducedamplitude}
\end{equation}
Here $\mathcal F_q$ is an unknown reduced amplitude containing the axial and pseudoscalar pieces, phases, mass-ratio functions and logarithms. Its evaluation requires the dimuon matching. Both the SM and new amplitudes can start at one loop, so their ratio retains the relative couplings, loop functions and flavour factors. This dimensionless amplitude ratio determines the sensitivity.

For illustration, in the axial-only, real-amplitude direction $\Delta C_{10}^{(q)}-\Delta C_{10}^{\prime(q)}=C_{10,\rm SM}^{(q)}\epsilon_R^{(6)}\mathcal F_q$. Using $C_{10,\rm SM}\simeq-4.1$ as a representative low-scale normalisation, a unit reduced coefficient corresponds to $|\delta P_q|=2.58\times10^{-4}$ and $|\Delta(C_{10}-C_{10}')|\simeq1.06\times10^{-3}$ at $5\,\mathrm{TeV}$. At $20\,\mathrm{TeV}$ these numbers are $1.62\times10^{-5}$ and $6.62\times10^{-5}$. These values follow from power counting with the specified unit reduced coefficient.

The $B_s$ branching-fraction error is approximately $8\%$ relative to its SM reference, so a representative one-standard-deviation amplitude scale is $|\delta P_s|\sim0.04$ for real interference with $\mathcal A_{\Delta\Gamma_s}^{\mu\mu}=+1$. Equation~\eqref{eq:reducedamplitude} then requires
\begin{equation}
 |\mathcal F_s|\sim(155,619,2476,9906,39624)
 \label{eq:requiredF}
\end{equation}
at $M_{W_R}=(5,10,20,40,80)\,\mathrm{TeV}$. The $B_d$ error probes a still larger amplitude change, of order $0.2$. Under the illustrative order-one assumption, the corresponding rate shifts lie below current experimental precision. The estimate places no upper bound on $|\mathcal F_q|$; the sensitivity of the full model remains undetermined pending the amplitude calculation.

The computed $\Delta F=2$ amplitudes give a more concrete yardstick than an assumed order-one coefficient. Define the absolute real-part ratio
\begin{equation}
 \mathcal R_s^{\rm mix}\equiv
 \frac{|\operatorname{Re}(M_{12,s}^{\rm NP}/M_{12,s}^{\rm SM})|}
 {\epsilon_R^{(6)}}.
 \label{eq:mixingyardstick}
\end{equation}
The five points give $\mathcal R_s^{\rm mix}=(57.7,67.7,77.8,87.9,97.9)$; its growth includes the logarithmic mass dependence of the mixing calculation. If the real interfering dimuon coefficient had this same magnitude, the linear rate-shift magnitudes would be $(2.98,0.875,0.251,0.0710,0.0198)\%$ under the rate assumptions above. Reaching eight percent instead requires coefficients larger by factors $(2.68,9.14,31.8,113,405)$. Thus even this enhanced yardstick falls short, with the smallest gap at the $5\,\mathrm{TeV}$ point already flagged by the conditional charged-search comparison. The comparison assumes equal magnitudes for the two reduced coefficients as a numerical reference. Their physical values depend on the distinct $\Delta B=2$ and dimuon operators, loop functions, chiral factors and phases.

A scalar enhancement must also be assessed with the actual vertices. Defining $h_q=M_{B_q}^2/[2m_\mu(m_b+m_q)]$, the pseudoscalar part is $\delta P_q|_P=h_q\Delta(C_P-C_P')/C_{10,\rm SM}$. At representative $m_b+m_s\simeq4.3\,\mathrm{GeV}$ and $M_{B_s}=5.37\,\mathrm{GeV}$ one has $h_s\simeq32$. Hence the same amplitude scale would require $|\Delta(C_P-C_P')|\sim5\times10^{-3}$ in our mass-independent scalar-operator convention, or a similar scalar coefficient for $|S_s|\sim0.04$. A pure scalar contribution enters the rate quadratically, giving rate-only sensitivity at $|S_s|\sim\sqrt{0.08}\simeq0.28$. The aligned reference has zero tree scalar FCNCs. The heavy scalar mass grows with $v_R$, and the light-heavy scalar mixing falls as $v_L/v_R$; the resulting scalar coefficient depends on the matched vertices. Neutral-vector contributions likewise require a flavour-changing fermion vertex. Gauge-independent matching determines whether these contributions enhance the dimuon amplitude.

\section{Reference spectrum, meson mixing and radiative decays}
\label{sec:reference}

We evaluate the selected texture for $M_{W_R}=5$, $10$, $20$, $40$ and $80\,\mathrm{TeV}$, with $M=v_R$, $v_L=246.220\,\mathrm{GeV}$, $g_L=g_R=0.652905$, $g_Y=0.357$ and $\kappa=0.05$. The quartic $\lambda$ is fixed by the exact light-scalar eigenvalue $m_h=125.25\,\mathrm{GeV}$. Gauge and Yukawa parameters are specified at a fixed reference scale. A multiscale parity-boundary fit requires their evolution and threshold matching. Quark running masses at $\mu=M_W$ and the central CKM matrix are taken from \texttt{flavio}~\cite{flavio}; charged-lepton pole masses are used at this order. The illustrative normal-ordering Dirac spectrum has $m_1=0.001\,\mathrm{eV}$, $\Delta m_{21}^2=7.42\times10^{-5}\,\mathrm{eV}^2$ and $\Delta m_{31}^2=2.517\times10^{-3}\,\mathrm{eV}^2$. Its sum is approximately $0.060\,\mathrm{eV}$.

For example, the $80\,\mathrm{TeV}$ reference has $v_R=245.059\,\mathrm{TeV}$, $M_{Z_2}=95.548\,\mathrm{TeV}$ and $M_H=126.886\,\mathrm{TeV}$. Its heavy up-type masses are approximately $(245.059,245.377,345.461)\,\mathrm{TeV}$, despite the degenerate bare mass $M$. The largest Yukawa singular value is $1.405$, $\lambda=0.13405$ and the light-heavy radial-scalar mixing angle is $1.87\times10^{-4}$. The nonzero CKM invariant, $J=3.05\times10^{-5}$, is retained. The reference CKM inputs include $|V_{us}|=0.2248$, $|V_{cb}|=0.04197$, $|V_{ub}|=0.00364$ and $\gamma=65.89^\circ$. The stated value of $V_{cb}$ defines the quoted $\epsilon_K$ result. The reference SM predictions are $\Delta M_s=1.14130\times10^{-11}\,\mathrm{GeV}$, $\Delta M_d=3.45509\times10^{-13}\,\mathrm{GeV}$, $\epsilon_K=1.95289\times10^{-3}$ and $\mathcal B(B\to X_s\gamma)=3.30969\times10^{-4}$. These values and the fixed CKM inputs define the preserved package snapshot used throughout the reference comparison.

For neutral-meson mixing we sum over all six internal quarks in the $W_LW_L$, $W_RW_R$ and $W_LW_R$ boxes, including the associated Goldstone terms. We subtract the three-generation SM left-handed amplitude before adding new coefficients to the SM prediction. The scalar LR operator has the particularly relevant matching
\begin{equation}
 \begin{aligned}
 H_{LR}^{qb}={}&\frac{g_L^2g_R^2}{64\pi^2M_{W_R}^2}
 \sum_{i,j}\lambda_i^{LR}\lambda_j^{RL}\sqrt{x_ix_j}\\
 &\times\big[(4+\beta x_ix_j)I_1(x_i,x_j;\beta)\\
 &\qquad-(1+\beta)I_2(x_i,x_j;\beta)\big]\\
 &\times(\bar qP_Lb)(\bar qP_Rb)+\mathrm{h.c.}
 \end{aligned}
 \label{eq:mixedbox}
\end{equation}
where $x_i=m_i^2/M_W^2$, $\beta=M_W^2/M_{W_R}^2$, $\lambda_i^{AB}=(\mathbb V_A)_{iq}^*(\mathbb V_B)_{ib}$ and $I_{1,2}$ are specified in \ref{app:kernels}. The same construction applies to kaons and, with internal down-type states and the appropriate conjugations, to charm. The mass-weighted identity $\sum_i m_i(\mathbb V_L)_{iq}^*(\mathbb V_R)_{ib}=0$ follows from the zero upper-left block of Eq.~\eqref{eq:commonfermion}. Retaining the heavy states preserves this identity and the required cancellations. We evaluate the box kernels of Ref.~\cite{Dcruz} with exact internal masses and fermion rotations.

The matched $\Delta F=2$ operators are evolved to hadronic scales with \texttt{wilson}, and their matrix elements are evaluated with \texttt{flavio}~\cite{wilson,flavio}. Table~\ref{tab:allflavour} reports all five points, including $B_d$ mixing and the absolute short-distance shift $\delta\Delta M_K=2\operatorname{Re}M_{12,K}^{\rm NP}$. A total kaon-mass prediction also requires the long-distance amplitude. The $\epsilon_K$ column retains the sign of the reference calculation, including its reversal at $5\,\mathrm{TeV}$. The charm amplitude is also available, but a charm fit would require long-distance nuisance amplitudes. The table records the predictions at the fixed reference inputs.

\begin{table*}[t]
\centering\small
\caption{Complete five-point reference results. Upper panel: total reference predictions for the first four observables and the new short-distance kaon mass shift; the radiative branching fraction marked $\gamma^{\rm dip}$ includes charged-vector dipoles only. Lower panel: dispersive-amplitude and kaon-CP shifts relative to the reference SM. The displayed digits resolve small shifts within this fixed-order calculation. The radiative column is restricted to the charged-vector dipoles.}
\label{tab:allflavour}
\begin{tabular}{rrrrrr}
\toprule
$M_{W_R}$ [TeV] & $\Delta M_s/10^{-11}$ [GeV] & $\Delta M_d/10^{-13}$ [GeV] & $\epsilon_K/10^{-3}$ & $\mathcal B_\gamma^{\rm dip}/10^{-4}$ & $\delta\Delta M_K/10^{-17}$ [GeV]\\
\midrule
5 & 1.124290 & 3.403556 & -1.327821 & 3.30978710 & -61.743949\\
10 & 1.136303 & 3.439954 & 1.007279 & 3.30971598 & -16.138227\\
20 & 1.139863 & 3.450742 & 1.685173 & 3.30969820 & -4.209720\\
40 & 1.140893 & 3.453860 & 1.878137 & 3.30969375 & -1.096191\\
80 & 1.141185 & 3.454745 & 1.932247 & 3.30969264 & -0.284986\\
\bottomrule
\end{tabular}
\medskip
\begin{tabular}{rrrrr}
\toprule
$M_{W_R}$ [TeV] & $\operatorname{Re}(M_{12,s}^{\rm NP}/M_{12,s}^{\rm SM})$ & $\operatorname{Re}(M_{12,d}^{\rm NP}/M_{12,d}^{\rm SM})$ & $\operatorname{Re}(M_{12,K}^{\rm NP}/M_{12,K}^{\rm SM})$ & $\delta\epsilon_K/\epsilon_{K,\rm SM}$\\
\midrule
5 & -0.0149022 & -0.0149144 & -0.2135560 & -1.6799253\\
10 & -0.0043763 & -0.0043799 & -0.0558202 & -0.4842116\\
20 & -0.0012566 & -0.0012576 & -0.0145615 & -0.1370883\\
40 & -0.0003547 & -0.0003550 & -0.0037919 & -0.0382791\\
80 & -0.0000988 & -0.0000989 & -0.0009858 & -0.0105714\\
\bottomrule
\end{tabular}
\end{table*}

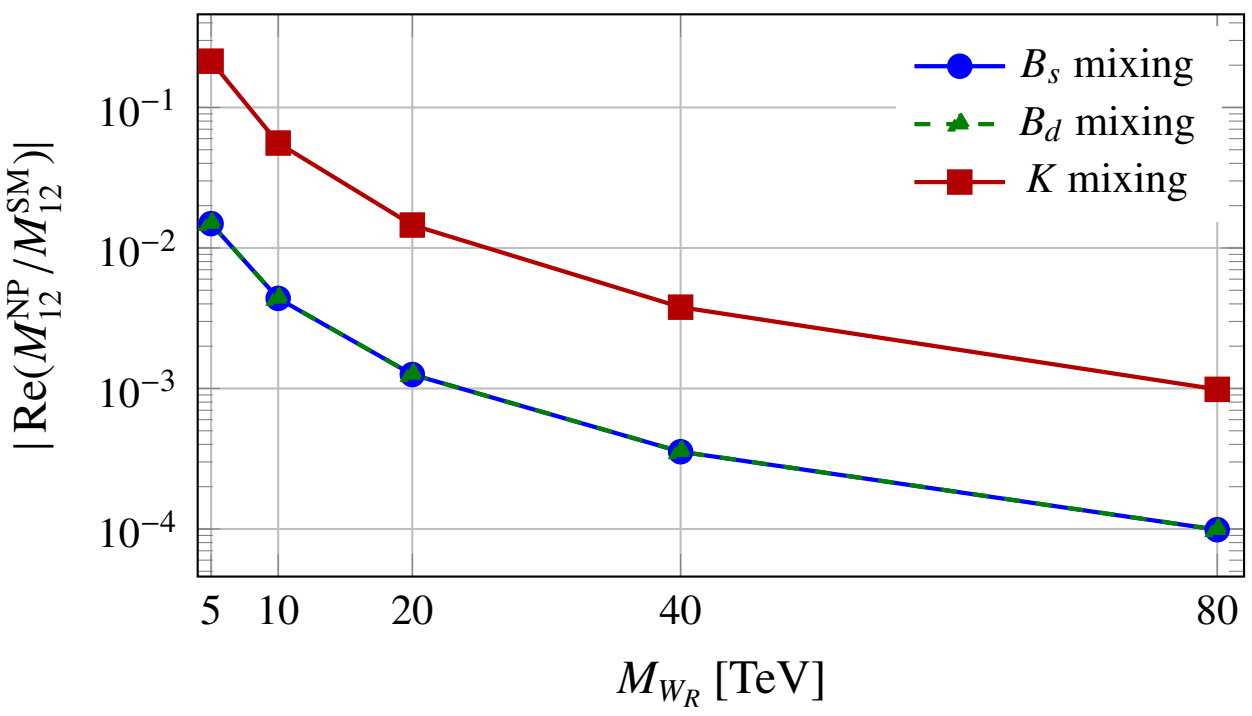
\begin{figure}[t]
\centering
\begin{tikzpicture}
\begin{semilogyaxis}[width=.96\columnwidth,height=5.3cm,
 xlabel={$M_{W_R}\,[\mathrm{TeV}]$},
 ylabel={$|\operatorname{Re}(M_{12}^{\mathrm{NP}}/M_{12}^{\mathrm{SM}})|$},
 xmin=4,xmax=82,xtick={5,10,20,40,80},grid=major,
 legend style={at={(.98,.98)},anchor=north east,draw=none,font=\footnotesize},
 tick label style={font=\footnotesize},label style={font=\small}]
\addplot[blue,thick,mark=*] coordinates {(5,0.0149021824507) (10,0.00437627357496) (20,0.00125655887284) (40,0.000354746146693) (80,9.88368656921e-05)};\addlegendentry{$B_s$ mixing}
\addplot[green!50!black,dashed,thick,mark=triangle*] coordinates {(5,.0149143730986) (10,.00437988089408) (20,.00125760050357) (40,.000355041494456) (80,.0000989194368231)};\addlegendentry{$B_d$ mixing}
\addplot[red!70!black,thick,mark=square*] coordinates {(5,0.213556016206) (10,0.0558202090995) (20,0.0145614827776) (40,0.00379187429977) (80,0.000985836825025)};\addlegendentry{$K$ mixing}
\end{semilogyaxis}
\end{tikzpicture}
\caption{Charged-vector reference contributions to dispersive meson mixing in the aligned spectrum with $M_F=v_R\mathbf1$. The $B_d$ and $B_s$ curves nearly coincide. The plotted real parts are negative; magnitudes are shown on a logarithmic scale. The points reproduce the existing reference scan, and the connecting lines guide the eye. The vertical axis gives the fractional dispersive-mixing contribution.}
\label{fig:flavourscan}
\end{figure}

The radiative entries retain only the matched charged-vector dipoles $C_{7,8}^{(\prime)}$. Additional amplitudes can increase or decrease the full rate through interference. The dipole-only column therefore supplies no conservative bound on the full model. In particular a radiatively generated $W_L$--$W_R$ insertion can acquire a top-to-bottom chiral enhancement. A quantitative size check is obtained from the quark-loop zero-momentum self-energy insertion
\begin{equation}
 \begin{aligned}
 \Pi_{LR}^{(q)}(0)={}&\frac{3g_Lg_R}{16\pi^2}
 \sum_{i,j=1}^{6}m_{u_i}m_{d_j}
 (\mathbb V_L)_{ij}(\mathbb V_R)^*_{ij}\\
 &\times B_0^{\rm fin}(0;m_{u_i}^2,m_{d_j}^2),\\
 \zeta_0&=\frac{|\Pi_{LR}^{(q)}(0)|}{M_{W_R}^2-M_W^2}.
 \end{aligned}
 \label{eq:zetaproxy}
\end{equation}
The finite function and cancellation of the constant part are specified in \ref{app:diagnostics}. The insertion measures the quark-loop self-energy at zero momentum. An on-shell mixing parameter requires momentum-dependent renormalisation. With the complete six-state rotations, $\zeta_0=(9.46\times10^{-7},2.76\times10^{-7},7.89\times10^{-8},2.22\times10^{-8},6.17\times10^{-9})$ at the five masses. The light-top estimate
\begin{equation}
 \delta C_7\sim\frac{g_R}{g_L}\zeta_0\frac{m_t}{m_b}
 \frac{V_{R,tb}}{V_{L,tb}}F_7^{LR}
 \label{eq:chiralradiative}
\end{equation}
contains $\zeta_0m_t/m_b$ between $5.62\times10^{-5}$ and $3.66\times10^{-7}$ before the mixing ratio and loop function. It can exceed the very small changes in the dipole-only column. Heavy-state terms, momentum dependence, four-quark mixing and matrix elements remain necessary. The rate comparison is therefore conditional on the amplitudes retained.

For a comparison scale, the standard effective low-energy dipole coefficient has $C_{7,\rm SM}^{\rm eff}(\mu_b)\simeq-0.30$~\cite{Buras}. Parameterise an illustrative low-scale contribution as $\delta C_7^{\rm eff}=X_7\mathcal L_7$, where $X_7=\zeta_0m_t/m_b$. The unknown $\mathcal L_7$ absorbs the right-handed mixing ratio, loop functions and evolution from matching to $\mu_b$; setting it to one fixes the normalisation for the following size comparison. For a real interfering shift in the SM-helicity coefficient, dipole dominance gives
\begin{equation}
 \frac{\delta\mathcal B_\gamma}{\mathcal B_{\gamma,\rm SM}}
 \simeq2\operatorname{Re}\frac{\delta C_7^{\rm eff}}{C_{7,\rm SM}^{\rm eff}}.
 \label{eq:radiativecomparison}
\end{equation}
A primed opposite-helicity shift enters quadratically in the massless strange-quark limit. With $|\mathcal L_7|=1$, the linear rate scales from the insertion are $0.0375\%$ at $5\,\mathrm{TeV}$ and $0.000244\%$ at $80\,\mathrm{TeV}$, compared with roughly $5\%$ experimental precision ($4.3\%$ in the archival input). An effective factor of order $130$ would be needed at the lowest mass for a five-percent linear shift. The computed dipole-only rates and this insertion estimate therefore suggest weak radiative sensitivity on the reference trajectory for modest factors. The full radiative sensitivity also depends on the additional amplitudes and any enhancement of their matched coefficients.

For transparency, \ref{app:historical} reproduces the historical four-observable Gaussian statistic with its sign, covariance, fixed inputs and per-observable decomposition. Its changes are positive and predominantly reflect a worsening of $\epsilon_K$ relative to an SM central value already below the data. The penalties are $+805.5$, $+96.0$, $+16.0$, $+3.58$ and $+0.920$ at the five masses. The correlated kaon-CP contributions alone account for approximately $793.5$, $92.0$, $14.81$, $3.24$ and $0.826$ of those changes. The $5\,\mathrm{TeV}$ point even reverses the signed reference $\epsilon_K$. Thus, on these fixed archival inputs, kaon CP violation strongly penalises the lower masses and retains a substantial preference against the $20$--$40\,\mathrm{TeV}$ reference range. The penalty is already much weaker at $40\,\mathrm{TeV}$ than at $20\,\mathrm{TeV}$. The statistical interpretation is conditional on these fixed archival inputs. Extracting a mass limit requires CKM profiling, updated data, the full radiative amplitude and likelihood calibration. Table~\ref{tab:allflavour} supplies the underlying observable predictions for that analysis.

\section{Charged and neutral resonance checks}
\label{sec:collider}
Light invisible Dirac partners give the charged-lepton plus missing-momentum signature $W_R\to\ell\nu_R$. For the reference rotations and closed heavy-fermion thresholds, the massless narrow-width calculation gives
\begin{equation}
 \Gamma_{W_R}/M_{W_R}=0.0296,\qquad
 \mathcal B(W_R\to e\nu_R,\mu\nu_R)=0.1909.
\end{equation}
Compared at equal mass with a sequential-SM $W'$ with fermionic SM-strength couplings, the minimum light-quark production weight times the branching-fraction ratio is $1.129$. It includes the right-handed doublet admixtures in production and decay. The published CMS reference excludes its sequential-SM state below $5.7\,\mathrm{TeV}$~\cite{CMSWprime}. If the two models have the same acceptance and a narrow-resonance rate interpretation, the reference $W_R$ rate is larger at fixed mass and the $5\,\mathrm{TeV}$ point is excluded by that comparison. This inherited rate bound assumes equal acceptance. A detector-level LR recast requires the spin-dependent acceptance, off-shell tails and signal modelling. The $5\,\mathrm{TeV}$ point fails the stated conditional rate comparison.

The neutral state is also checked at the level of its actual fermionic couplings. The ordinary-muon limit gives $g_{\mu L}=0.1165$ and $g_{\mu R}=-0.1567$. Including the reference light-fermion rotations and invisible Dirac channels, but keeping only massless fermionic widths, gives $\Gamma_{Z_2}/M_{Z_2}=0.0244$ and $\mathcal B(Z_2\to e^+e^-,\mu^+\mu^-)=0.0415$. For each initial quark define
\begin{equation}
 c_q=[|g_{qL}|^2+|g_{qR}|^2]\mathcal B_{e\mu},\qquad
 \frac{c_u}{c_u^{\rm SSM}}=0.478,\quad
 \frac{c_d}{c_d^{\rm SSM}}=0.650.
 \label{eq:neutralrates}
\end{equation}
Here the SSM neutral reference uses SM $Z$ chiral couplings and ordinary fermionic channels. These ratios are essentially constant along the scan. Each initial-state ratio compares the two models at the same resonance mass, where the common parton luminosity cancels. The lowest scanned neutral mass is $5.97\,\mathrm{TeV}$, already above the $5.1\,\mathrm{TeV}$ SSM reference exclusion in the cited ATLAS search~\cite{ATLASdilepton}; the scanned neutral masses therefore begin above that SSM threshold. An LR neutral-mass bound requires the absolute cross sections, bosonic widths, interference, fiducial acceptance and observed limit curve.

\section{Thermal history and radiative strong CP}
Light Dirac neutrinos introduce an independent restriction on the common spectrum. Assume that all three right-handed species were initially thermal, that they decouple after the heavy LR states have left the bath, and that no subsequent exotic entropy release dilutes them. Entropy conservation then gives~\cite{Luo}
\begin{equation}
 \Delta N_{\rm eff}=3\left[\frac{g_{*s}(T_{\nu_L\rm dec})}
 {g_{*s}(T_{\nu_R\rm dec})}\right]^{4/3}
 \geq3\left(\frac{10.75}{106.75}\right)^{4/3}=0.1405.
 \label{eq:nefffloor}
\end{equation}
The inequality gives the entropy-counting lower bound on the radiation contribution. Any later decoupling below SM mass thresholds increases the contribution. The specified initially thermal history fixes this minimum abundance, including at large right-handed mass.

The mass dependence can be estimated with a charged-current-only freeze-out proxy. The estimate omits $Z_R$-mediated scattering and annihilation from the collision rate. It follows from $G_R=g_R^2/(4\sqrt2M_{W_R}^2)$ and $\Gamma\sim G_R^2T^5$, giving the scaling estimate~\cite{Luo}
\begin{equation}
 T_R^{\rm proxy}=2\,\mathrm{MeV}\,(G_R/G_F)^{-2/3}
 \propto M_{W_R}^{4/3}.
 \label{eq:thermalproxy}
\end{equation}
The proxy represents the charged collision channels by one effective coefficient. At the five scan masses it gives $T_R^{\rm proxy}=(0.49,1.24,3.13,7.89,19.87)\,\mathrm{GeV}$. In an illustrative free quark-gluon gas with finite charm, bottom, muon and tau masses, the corresponding entropy factors give $\Delta N_{\rm eff}^{\rm proxy}=(0.254,0.215,0.193,0.188,0.187)$. Near the lowest temperature QCD interactions make a free-gas treatment especially crude; neutral scattering and channel multiplicities also change the decoupling normalisation. These values illustrate the mass dependence and the increase above the $0.1405$ floor. A precise thermal prediction requires the full charged and neutral collision integrals. The scale dependence and numerical integration defining this proxy are given in \ref{app:diagnostics}.

The dataset dependence matters. The free-streaming fits in Ref.~\cite{ACT} give $N_{\rm eff}=2.86\pm0.13$ for P-ACT-LB, $2.87\pm0.13$ after adding deuterium, and $2.89\pm0.11$ after adding deuterium and helium. The floor $N_{\rm eff}\geq3.1845$ lies respectively $2.50$, $2.42$ and $2.68$ Gaussian standard deviations above those central values; the SM reference $3.044$ lies $1.42$, $1.34$ and $1.40$ above them. The quoted Gaussian pulls describe one-dimensional comparisons for overlapping datasets. The same study quotes $3.06\pm0.17$ for Planck-LB, for which the floor is only $0.73$ standard deviations high. The inferred tension depends on the dataset, cosmological model and thermal history.

Under the initially thermal, undiluted branch, every scan point has this conditional radiation tension, and the freeze-out proxy makes it stronger. A global assessment requires combining this thermal test with the flavour and collider likelihoods. Low reheating and entropy dilution introduce further thermal-history parameters whose values would need to be specified and fitted.

For strong CP, the real doublet vacuum and Hermitian singlet masses give
\begin{equation}
 \det\mathcal M_f=(-1)^3\left(\frac{v_Lv_R}{2}\right)^3
 |\det Y_f|^2,\qquad
 \arg\det(\mathcal M_u\mathcal M_d)=0.
\end{equation}
The established parity-based determinant cancellation~\cite{Hisano,Banno} therefore holds at tree level with a nonzero CKM phase. A neutron-EDM prediction also requires the radiative mass phases and dipole amplitudes. To examine the leading mass-phase structures, define $(\mathsf H_f^a)_{ij}=(Y_f)_{ia}(Y_f)^*_{ja}$ and $(\mathsf R_f^a)_{ij}=(Y_f)_{ia}(Y_f^{-1})_{aj}$ without summing $a$. The leading structures discussed in the recent universal-seesaw analysis~\cite{Banno} include
\begin{equation}
 \operatorname{Im}\operatorname{Tr}([\mathsf H_u^1,\mathsf H_u^2]\mathsf H_u^3),\qquad
 \operatorname{Im}\operatorname{Tr}(\mathsf H_u^a\mathsf H_u^b\mathsf R_d^c),\qquad
 \operatorname{Im}\operatorname{Tr}(\mathsf H_d^a\mathsf H_d^b\mathsf R_u^c).
\end{equation}
In Eq.~\eqref{eq:alignment}, $\mathsf H_f^a=y_{fa}^2P_f^a$ and $\mathsf R_f^a=P_f^a$, with orthogonal same-sector Hermitian projectors. The commutators vanish; the mixed traces reduce to real traces of two projectors. These leading invariants therefore vanish analytically, and the calculation checks the cancellation with the CKM phase retained. Generic singlet flavour textures can give nonzero invariants. The radiative calculation also requires heavy-quark threshold matching.

For degenerate bare singlet masses the surviving CKM contribution can be organised through the much smaller Jarlskog structure. The scaling estimate of Ref.~\cite{Hisano} is
\begin{equation}
 \delta\bar\theta_{\rm CKM}\sim
 \frac{Jf_{\rm CKM}}{(16\pi^2)^2}\frac{(v_R/\sqrt2)^2}{M^2}
 \frac{\displaystyle\prod_{i<j}(m_{u_j}-m_{u_i})
 \prod_{i<j}(m_{d_j}-m_{d_i})}{(v_L/\sqrt2)^6}.
\end{equation}
Here $f_{\rm CKM}$ is the dimensionless loop function left unevaluated in this degenerate-mass scaling estimate. Its value requires threshold matching at the exact spectrum. The reproducible statement is $\delta\bar\theta_{\rm CKM}\sim(1.75\times10^{-19})f_{\rm CKM}$ for $M=v_R$: the displayed coefficient is obtained by setting $f_{\rm CKM}=1$. The products use pairwise differences of the running masses $(m_u,m_c,m_t)=(0.00125725,0.638379,172.438)\,\mathrm{GeV}$ and $(m_d,m_s,m_b)=(0.00272113,0.0538232,2.90294)\,\mathrm{GeV}$ at $\mu=M_W$. The coefficient is constant along this fixed-input trajectory; the loop function and neutron-EDM prediction remain unevaluated. The boundary-scale evolution of the flavour limit is discussed below. The experimental requirement remains $|d_n|<1.8\times10^{-26}\,e\,\mathrm{cm}$ at 90\% confidence~\cite{nEDM}. A final likelihood must include radiative mass phases, quark electric and chromoelectric dipoles, the relevant thresholds and hadronic uncertainties.

\section{Boundary-scale stability of the flavour limit}
\label{sec:stability}
We impose singlet degeneracy at the matching scale. Yukawa interactions drive departures from that boundary condition under scale evolution. At one loop a singlet mass has flavour structures of the form $M_FY_f^\dagger Y_f+Y_f^\dagger Y_fM_F$, up to calculable numerical coefficients and gauge terms proportional to $M_F$. A top Yukawa of $1.405$ permits a fractional nonuniversal correction of order $0.0125\ln(\Lambda/\mu)$ in the up-singlet sector. This parametric estimate indicates the evolution needed to relate a high-scale degenerate boundary to the reference matching scale.

An off-diagonal down-sector FCNC requires misalignment between the singlet mass matrix and the Yukawa basis. At the degenerate boundary, the one-loop down-singlet structure $Y_d^\dagger Y_d$ is diagonal in the right-singular basis of $Y_d$. A top spurion enters through the coupled Yukawa evolution, or higher-order structures such as $Y_d^\dagger Y_uY_u^\dagger Y_d$. The coupled RGEs determine the leading-log coefficient and the order at which an off-diagonal $M_D$ appears. At a fixed scale with $M_F\propto\mathbf1$, unitary Yukawa re-diagonalisation preserves the neutral FCNC cancellation.

For a quantitative envelope, consider an off-diagonal singlet perturbation in the down Yukawa basis of size
\begin{equation}
 \begin{aligned}
 |\eta_{sb}|&\equiv|\delta M_{D,sb}|/M\\
 &\sim\frac{y_t^2|V_{ts}^*V_{tb}|}{16\pi^2}L_\Lambda
 =5.15\times10^{-4}L_\Lambda,\\
 L_\Lambda&=|\ln(\Lambda/\mu)|.
 \end{aligned}
 \label{eq:etaenvelope}
\end{equation}
This envelope allows a perturbation as large as a one-loop flavour correction. Coupled running may introduce further down-Yukawa or loop suppression. Writing $g_Z=\sqrt{g_L^2+g_Y^2}$ and $g_{RB}=\sqrt{g_R^2+g_B^2}$, if $s_{Lq},s_{Rq}$ denote the ordinary-singlet mixing angles obtained from the exact two-state blocks, the neutral-vertex power counting is
\begin{equation}
 \begin{aligned}
 |\delta g_{sb}^{Z_1,L}|&\sim\frac{g_Z}{2}s_{Ls}s_{Lb}|\eta_{sb}|,\\
 |\delta g_{sb}^{Z_R,R}|&\sim\frac{g_R^2}{2g_{RB}}s_{Rs}s_{Rb}|\eta_{sb}|.
 \end{aligned}
 \label{eq:fcncenvelope}
\end{equation}
At $5\,\mathrm{TeV}$ these scales are $7.9\times10^{-11}L_\Lambda$ and $2.3\times10^{-7}L_\Lambda$. The first falls approximately as $M_{W_R}^{-2}$, while the second is nearly constant because right-handed admixtures approach fixed values on $M=v_R$; its exchange coefficient still falls as $M_{Z_R}^{-2}$. These vertex scales quantify the response to the assumed singlet perturbation. Matching them to dimuon coefficients requires specifying the ultraviolet boundary conditions and evolving the coupled flavour matrices. The tree determinant and CP-odd invariant cancellations likewise require the radiatively corrected flavour matrices before becoming an EDM prediction.

\section{Scope of a flavour likelihood}
A quantitative interpretation of the dimuon data requires the penguin, box and scalar matching of the chosen theory, including the complete fermion rotations. Charged-current loops contribute to the dimuon amplitude in the aligned model. The effective Hamiltonian identifies the required $C_{10}^{(\prime)}$, $C_S^{(\prime)}$ and $C_P^{(\prime)}$ shifts and their phase combinations. Their full one-loop matching remains to be evaluated.

A completed flavour likelihood must combine the $B_d$ and time-integrated $B_s$ dimuon branching fractions, the effective lifetime, neutral-meson mixing and the complete radiative amplitude. It must retain correlations, lifetime-dependent efficiencies, CKM and hadronic nuisance parameters. Each experimental input enters once, through either the combined measurement or its correlated constituents. The signed historical reproduction in \ref{app:historical} documents the fixed-input diagnostic and its covariance. Collider statements in Sec.~\ref{sec:collider} are explicitly conditional rate comparisons. The thermal branch is subject to the dataset-dependent radiation tension quantified above.

\section{Conclusions}
The degenerate-singlet limit of the parity-related doublet universal seesaw has no light-light tree neutral flavour transition, independently of the basis used to express its Yukawa matrices. Charged-vector mixing boxes nevertheless survive. The complete five-point reference results show a substantially larger fractional effect in kaon CP violation than in $B_{d,s}$ dispersive mixing. The axial selection rule identifies the separate amplitude combinations measured by the dimuon channels.

The numerical comparisons make the hierarchy of probes explicit on the fixed-ratio decoupling trajectory. The mixing reduced-coefficient yardstick grows from $57.7$ to $97.9$, whereas an eight-percent dimuon shift requires $155$ to $39624$. Assigning equal magnitudes to the two reduced coefficients gives the comparison rate scales $2.98\%$ down to $0.0198\%$. The archival kaon-CP comparison penalises the $20$ and $40\,\mathrm{TeV}$ points by total changes $\Delta\chi^2=+16.0$ and $+3.58$, dominated by $\epsilon_K$; lower masses are much more strongly disfavoured within that fixed-input diagnostic. The result locates the sensitivity of the fixed-input diagnostic in the $20$--$40\,\mathrm{TeV}$ range. For a unit effective loop-and-evolution factor, the radiative insertion scale remains below $0.04\%$, far below the roughly $5\%$ experimental rate precision. The full radiative and dimuon sensitivity follows from their matched amplitudes. The charged-search rate comparison already flags the $5\,\mathrm{TeV}$ point under equal-acceptance assumptions, while the scanned neutral masses lie above the cited SSM exclusion threshold. The thermal, undiluted Dirac branch is in dataset-dependent radiation tension; the scalar hierarchy also requires the explicit cancellation in Eq.~\eqref{eq:vacuumtuning}.

The operator analysis and numerical reference results connect the flavour tests to a common spectrum and explicit boundary conditions. A global fit of that spectrum requires full dimuon and radiative matching, collider recasts, coupled running and thermal evolution, together with updated correlated experimental inputs.

\appendix
\section{Symmetry breaking and the gauge spectrum}
\label{app:lrbreaking}
The two scalar doublets have components $\chi_A=(\chi_A^+,\chi_A^0)^T$, with $\langle\chi_A^0\rangle=v_A/\sqrt2$. The lower component of $\chi_R$ has $T_R^3=-1/2$ and $t_{BL}=1/2$. Its expectation value is therefore neutral under $Y=T_R^3+t_{BL}$ and breaks the right-handed gauge factors to hypercharge. The subsequent left-handed expectation value leaves $Q_{\rm em}=T_L^3+Y$ unbroken:
\begin{equation}
 SU(2)_L\times SU(2)_R\times U(1)_{B-L}
 \xrightarrow{\ v_R\ }SU(2)_L\times U(1)_Y
 \xrightarrow{\ v_L\ }U(1)_{\rm em}.
 \label{eq:appbreaking}
\end{equation}
This ordered description is useful for $v_R\gg v_L$; the mass matrix in Eq.~\eqref{eq:exactneutral} retains both expectation values simultaneously.

The vector masses follow from $|D_\mu\chi_L|^2+|D_\mu\chi_R|^2$. Because no scalar transforms as a doublet under both weak groups, there is no tree charged mass term connecting $W_L$ to $W_R$. Consequently
\begin{equation}
 M_{W_L}^2=\frac{g_L^2v_L^2}{4},\qquad
 M_{W_R}^2=\frac{g_R^2v_R^2}{4}.
\end{equation}
This statement is specific to the doublet-only scalar content. A bidoublet $\Phi\sim(1,2,2,0)$ permits both direct ordinary-fermion Yukawa masses and a tree charged-vector mixing term; its inclusion therefore changes the scalar and fermion mass sectors.

To leading order in $v_L/v_R$, write $g_{RB}=\sqrt{g_R^2+g_B^2}$. The orthogonal neutral combinations are
\begin{equation}
 B_Y=\frac{g_BW_R^3+g_RX}{g_{RB}},\qquad
 Z_R=\frac{g_RW_R^3-g_BX}{g_{RB}},\qquad
 g_Y=\frac{g_Rg_B}{g_{RB}}.
\end{equation}
Substitution into the original interaction yields the hypercharge vertex $g_YYB_Y$ and the heavy-neutral vertex
\begin{equation}
 Z_{R\mu}\bar f\gamma^\mu
 \frac{g_R^2T_R^3-g_B^2t_{BL}}{g_{RB}}f.
\end{equation}
The second expression is understood separately on the two chiral representations. It gives different charged-lepton couplings even though $B-L$ itself is vectorial. At this order,
\begin{equation}
 M_{Z_R}^2=\frac{g_{RB}^2v_R^2}{4},\qquad
 \frac{M_{Z_R}}{M_{W_R}}=\frac{g_{RB}}{g_R}.
\end{equation}
The two heavy masses are therefore correlated. After the left-handed breaking, the massless photon obeys
\begin{equation}
 A_\mu=\frac{e}{g_L}W_{L\mu}^3+
 \frac{e}{g_R}W_{R\mu}^3+\frac{e}{g_B}X_\mu,\qquad
 \frac1{e^2}=\frac1{g_L^2}+\frac1{g_R^2}+\frac1{g_B^2}.
\end{equation}
The normalisation assumes canonical gauge kinetic terms. Finite $v_L/v_R$ mixes the two massive neutral states; low-energy matching then uses the exact eigenstates $Z_1,Z_2$.

The scalar spectrum also follows from the specified potential. Two complex doublets contain eight real fields. Six become Goldstone modes of the six massive gauge bosons, leaving two neutral CP-even scalars. Writing $\chi_A^0=(v_A+h_A+ia_A)/\sqrt2$ and choosing $h=\cos\theta_h h_L-\sin\theta_h h_R$, $H=\sin\theta_h h_L+\cos\theta_h h_R$, one obtains
\begin{align}
 m_{h,H}^2&=\lambda(v_L^2+v_R^2)
 \mp\sqrt{\lambda^2(v_R^2-v_L^2)^2+\kappa^2v_L^2v_R^2},\\
 \tan(2\theta_h)&=\frac{\kappa v_Lv_R}{\lambda(v_R^2-v_L^2)}.
\end{align}
The angle is chosen on the branch with $h$ predominantly $h_L$ for a hierarchical vacuum. The physical scalar spectrum consists of these two neutral states. Charged scalar components supply the Goldstone modes that enter covariant-gauge loop calculations.

\section{Universal seesaw masses and flavour rotations}
\label{app:lrmasses}
In the doublet scalar theory, the universal seesaw generates ordinary-fermion Dirac masses through vector-like weak singlets. For example, $\bar Q_LY_d\chi_LD_R$, $\bar Q_RY_d\chi_RD_L$ and $\bar D_LM_DD_R$ connect the light left and right quarks through the heavy sector. The adjective vector-like means that $F_L$ and $F_R$ have the same gauge representation, allowing a bare Dirac mass. The physical weak couplings depend on the resulting light-heavy rotations~\cite{Dcruz}.

For any charge sector, set $X_f=Y_fv_L/\sqrt2$ and $Z_f=Y_f^\dagger v_R/\sqrt2$. The mass matrix is $\mathcal M_f=(\begin{smallmatrix}0&X_f\\Z_f&M_F\end{smallmatrix})$. When both $X_fM_F^{-1}$ and $M_F^{-1}Z_f$ are small in norm, eliminating the heavy fields gives the leading light matrix
\begin{equation}
 m_f^{\rm eff}=-X_fM_F^{-1}Z_f
 =-\frac{v_Lv_R}{2}Y_fM_F^{-1}Y_f^\dagger.
 \label{eq:appseesaw}
\end{equation}
Its dimension is a mass. The expansion is controlled by the light-heavy mixing matrices. In particular, it can fail in the top sector of the reference spectrum.

The aligned texture reduces each generation to a real block $\mathcal M_i=(\begin{smallmatrix}0&a_i\\b_i&M\end{smallmatrix})$. The physical masses are the positive singular values of this generally nonsymmetric matrix. Defining $T_i=M^2+a_i^2+b_i^2$, they satisfy
\begin{equation}
 m_{i,\ell/h}^2=\frac12\left[T_i\mp\sqrt{T_i^2-4a_i^2b_i^2}\right],
 \qquad m_{i,\ell}m_{i,h}=|a_ib_i|.
 \label{eq:appsingular}
\end{equation}
For $a_i\ll\sqrt{M^2+b_i^2}$, one finds $m_{i,\ell}\simeq|a_ib_i|/\sqrt{M^2+b_i^2}$ and $m_{i,h}\simeq\sqrt{M^2+b_i^2}$. Only with the further condition $b_i\ll M$ does this become the familiar $|a_ib_i|/M$. The generation-dependent Yukawa couplings therefore split the physical heavy masses even for degenerate bare singlet masses.

The exact rotations $U_{fL}^\dagger\mathcal M_fU_{fR}=\widehat m_f$ must be applied to interaction generators as well as to mass terms. Since only the ordinary weak doublets carry charged weak generators, the complete quark and lepton vertices contain
\begin{equation}
 \mathbb V_A=U_{uA}^\dagger\Pi U_{dA},\qquad
 \mathbb K_A=U_{eA}^\dagger\Pi U_{\nu A},\qquad
 \Pi=\operatorname{diag}(\mathbf1_3,\mathbf0_3).
\end{equation}
Their light blocks are generally nonunitary because part of the doublet norm resides in heavy states. Neutral vertices instead contain $U_{fA}^\dagger T_AU_{fA}$. A generator proportional to the identity remains diagonal under any unitary rotation, but the ordinary doublets and the extra singlets have different weak generators. Generic light-heavy mixing can therefore induce flavour-changing neutral couplings. The aligned matrices selected in the main text make the light-light neutral vertices diagonal; this cancellation follows from the degenerate-singlet boundary condition at the stated scale.

The same construction can generate Dirac neutrino masses, but a gauge-singlet $N$ also admits Majorana terms unless an additional symmetry forbids them. The assumed lepton triality assigns charge one modulo three to each lepton and to $E,N$, and zero to scalars. A Dirac bilinear has net charge zero whereas a Majorana bilinear has charge two. Within the stated theory this preserves the Dirac spectrum. The light right-handed neutrinos then remain kinematically available in muon decay, even when their charged mediator is heavy. A triplet model with heavy Majorana states would have different final states, loop sums and direct-search constraints.

\section{Pseudoscalar-meson selection of the effective operators}
The vacuum-to-meson matrix element can be written
\begin{equation}
 \langle0|\bar q\gamma^\alpha\gamma_5b|\bar B_q(p)\rangle=if_{B_q}p^\alpha,
 \qquad \langle0|\bar q\gamma^\alpha b|\bar B_q(p)\rangle=0.
\end{equation}
For equal on-shell muon masses, with $p=p_-+p_+$, the Dirac equations give
\begin{equation}
 \begin{aligned}
 p_\alpha\bar u(p_-)\gamma^\alpha v(p_+)&=0,\\
 p_\alpha\bar u(p_-)\gamma^\alpha\gamma_5v(p_+)
 &=2m_\mu\bar u(p_-)\gamma_5v(p_+).
 \end{aligned}
\end{equation}
The $Q_9$ two-body matrix element therefore vanishes, while $Q_{10}$ contributes through $C_{10}-C_{10}'$ with helicity suppression. Scalar and pseudoscalar operators contribute through the differences included in $S_q$ and $P_q$. A comparison with semileptonic angular observables uses their distinct projections of the same matched Wilson coefficients.
\section{Loop kernels and matching conventions}
\label{app:kernels}
For the common-spectrum calculation it is useful to evaluate the box functions as convergent Euclidean integrals,
\begin{align}
 I_1(x,y;\beta)&=-\int_0^\infty\frac{t\,dt}{(t+x)(t+y)(t+1)(1+\beta t)},\\
 I_2(x,y;\beta)&=\int_0^\infty\frac{t^2\,dt}{(t+x)(t+y)(t+1)(1+\beta t)}.
\end{align}
These representations also apply when the internal masses coincide. The same-chirality Hamiltonian is
\begin{equation}
 \begin{aligned}
 H_{AA}^{qb}={}&\frac{g_A^4}{128\pi^2M_A^2}
 \sum_{i,j}\lambda_i^{AA}\lambda_j^{AA}\\
 &\times\bigg[\left(1+\frac{z_iz_j}{4}\right)I_2(z_i,z_j;1)\\
 &\qquad-2z_iz_jI_1(z_i,z_j;1)\bigg]O_{AA}+\mathrm{h.c.}
 \end{aligned}
\end{equation}
where $z_i=m_i^2/M_A^2$. The kernel reduces to the SM Inami--Lim function after GIM subtraction. Evolution uses the WCxf/\texttt{flavio} convention $\mathcal H=-\sum_iC_iO_i$; its dimensionful mixing coefficients therefore carry the opposite sign from the Hamiltonian coefficients displayed here.

For the dipole contribution the fixed-order matching uses the standard functions~\cite{Buras}
\begin{align}
 F_7(x)&=\frac{x(7-5x-8x^2)}{24(x-1)^3}
 +\frac{x^2(3x-2)\log x}{4(x-1)^4},\\
 F_8(x)&=\frac{x(2+5x-x^2)}{8(x-1)^3}
 -\frac{3x^2\log x}{4(x-1)^4}.
\end{align}
With $\lambda_t=V_{tq}^*V_{tb}$, the new dipole coefficients are
\begin{align}
 \Delta C_{7,8}&=\frac1{\lambda_t}\sum_{i=1}^6\lambda_i^{LL}
 F_{7,8}(m_i^2/M_W^2)-C_{7,8}^{\rm SM,LO},\\
 C_{7,8}'&=\left(\frac{g_R}{g_L}\right)^2\frac\beta{\lambda_t}
 \sum_{i=1}^6\lambda_i^{RR}F_{7,8}(m_i^2/M_{W_R}^2).
\end{align}
Charged-current four-quark Hamiltonians have coefficients $g_A^2(\mathbb V_A)_{iq}^*(\mathbb V_A)_{ib}/(2M_A^2)$ multiplying $(\bar q\gamma_\mu P_Au_i)(\bar u_i\gamma^\mu P_Ab)$. They are recorded separately from the dipole-only prediction. The broken-phase fixed-order amplitudes are supplied at $\mu=M_W$ and evolved below it; resumming the logarithms between the heavy thresholds and electroweak matching requires a multiscale UV-to-SMEFT calculation.

\section{Definition of the additional size diagnostics}
\label{app:diagnostics}
The estimates in the main text are separated from full matching predictions. For the quark-loop insertion in Eq.~\eqref{eq:zetaproxy}, one may choose
\begin{equation}
 B_0^{\rm fin}(0;x,y)=1-\frac{x\ln(x/\mu_0^2)-y\ln(y/\mu_0^2)}{x-y},
\end{equation}
with the continuous equal-mass limit $-\ln(x/\mu_0^2)$. The complete mass-weighted sum multiplying a constant is zero:
\begin{equation}
 \sum_{i,j}m_{u_i}m_{d_j}(\mathbb V_L)_{ij}(\mathbb V_R)^*_{ij}=0.
\end{equation}
It follows from the vanishing ordinary-doublet blocks of $\mathcal M_u$ and $\mathcal M_d$, or by tracing their product with the projectors. Both the ultraviolet constant and the arbitrary scale $\mu_0$ therefore cancel in this six-state fermionic sum. Omitting the heavy partners would spoil that cancellation. The quoted $\zeta_0$ uses fixed reference masses and is a zero-momentum insertion magnitude; an on-shell mixing parameter and its use in $b\to s\gamma$ require the appropriate momentum-dependent matching and counterterms.

For the neutral collider-rate check, a massless fermionic partial width is
\begin{equation}
 \frac{\Gamma(Z_2\to f\bar f)}{M_{Z_2}}
 =\frac{N_c^f}{24\pi}(|g_{fL}|^2+|g_{fR}|^2).
\end{equation}
We sum over the three light generations, including both chiralities of the Dirac neutrinos, with couplings obtained from the same six-state rotations. The width comparison retains the fermionic channels. The coefficients $c_q$ of Eq.~\eqref{eq:neutralrates} specify a parton-luminosity-weighted narrow-resonance rate; the fiducial event rate additionally requires detector efficiencies. The charged calculation analogously sums the light $\mathbb V_R$ and $\mathbb K_R$ norms, includes colour, and compares their production weights and $e+\mu$ branching fraction with the SSM reference at the same mass.

For the illustrative thermal proxy, the radiation bath includes photons, gluons, three active neutrinos, charged leptons and five quark flavours. For a free fermion of multiplicity $d_f$, set $z_f=m_f/T$, $E=\sqrt{x^2+z_f^2}$ and define
\begin{align}
 \frac{\rho_f}{T^4}&=\frac{d_f}{2\pi^2}\int_0^\infty\frac{x^2E\,dx}{e^E+1},\\
 \frac{p_f}{T^4}&=\frac{d_f}{6\pi^2}\int_0^\infty\frac{x^4\,dx}{E(e^E+1)},\\
 g_{*s}(T)&=\frac{45}{2\pi^2T^4}(\rho+p).
\end{align}
The illustrative threshold masses are $(m_u,m_d,m_s,m_c,m_b)=(0.0022,0.0047,0.093,1.27,4.18)\,\mathrm{GeV}$, with the charged-lepton masses of the reference inputs. These fixed threshold masses specify the illustrative free-gas entropy integrals. Evaluating these integrals at Eq.~\eqref{eq:thermalproxy} supplies the reported mass-dependent entropy proxy. A precision result requires the charged and neutral collision operators and a QCD equation of state, particularly near the lowest proxy temperature. The entropy floor follows directly from the maximal available SM entropy factor.

\begin{table*}[t]
\centering\small
\caption{Signed archival Gaussian comparison. The four $d_i$ are the correlated attributions defined in the text. All rows have positive total changes; the radiative entry retains charged-vector dipoles only.}
\label{tab:signeddiagnostic}
\begin{tabular}{rrrrrrr}
\toprule
$M_{W_R}$ [TeV] & $d_{\Delta M_s}$ & $d_{\Delta M_d}$ & $d_{\epsilon_K}$ & $d_{\mathcal B_\gamma^{\rm dip}}$ & $\Delta\chi^2$ & $\chi^2$\\
\midrule
5 & 6.14985 & 6.01544 & 793.52110 & -0.20884 & +805.47755 & 814.55826\\
10 & 1.35267 & 2.76481 & 91.98558 & -0.06007 & +96.04299 & 105.12370\\
20 & 0.34831 & 0.86859 & 14.80528 & -0.01700 & +16.00518 & 25.08589\\
40 & 0.09460 & 0.24931 & 3.24092 & -0.00474 & +3.58008 & 12.66079\\
80 & 0.02595 & 0.06935 & 0.82586 & -0.00131 & +0.91984 & 10.00055\\
\bottomrule
\end{tabular}
\end{table*}

\section{Signed reproduction of the historical Gaussian statistic}
\label{app:historical}
This appendix documents the inputs, covariance and signed contributions of the historical four-observable diagnostic. In the order $\mathbf O=(\Delta M_s,\Delta M_d,\epsilon_K,\mathcal B_\gamma^{\rm dip})$, let $\mathbf r(\Theta)=\mathbf O(\Theta)-\mathbf O_{\rm exp}$. The fixed-covariance comparison is
\begin{equation}
 \Delta\chi^2=\mathbf r^T\Sigma^{-1}\mathbf r
 -\mathbf r_{\rm SM}^T\Sigma^{-1}\mathbf r_{\rm SM}.
 \label{eq:signeddiagnostic}
\end{equation}
The experimental central values are $(1.1688\times10^{-11}\,\mathrm{GeV},3.334\times10^{-13}\,\mathrm{GeV},2.228\times10^{-3},3.27\times10^{-4})$, with errors $(1.40\times10^{-14}\,\mathrm{GeV},1.30\times10^{-15}\,\mathrm{GeV},1.10\times10^{-5},1.40\times10^{-5})$. They come from the package's HFAG 2015 mixing entries, PDG kaon-CP entry and 2017 radiative average. The theory covariance is the saved 300-draw SM reference, added to the diagonal experimental covariance. Its resulting total marginal standard deviations are $(4.76850\times10^{-13}\,\mathrm{GeV},2.20309\times10^{-14}\,\mathrm{GeV},1.57820\times10^{-4},2.56199\times10^{-5})$, with correlation matrix
\begin{equation}
 \rho=\begin{pmatrix}
 1&0.578538&0.220024&0.028733\\
 0.578538&1&0.577356&0.070065\\
 0.220024&0.577356&1&0.000347\\
 0.028733&0.070065&0.000347&1
 \end{pmatrix}.
\end{equation}
The central SM value is $\chi^2_{\rm SM}=9.08071$. The four observables are evaluated at fixed parameter points and fixed covariance. Interpreting this quadratic form as a global goodness-of-fit statistic or a mass limit requires updated inputs, a complete radiative amplitude, nuisance profiling and statistical calibration.

Table~\ref{tab:signeddiagnostic} exposes the signed changes and a defined correlated attribution,
\begin{equation}
 d_i=r_i(\Sigma^{-1}\mathbf r)_i
 -r_{{\rm SM},i}(\Sigma^{-1}\mathbf r_{\rm SM})_i,
 \qquad\sum_i d_i=\Delta\chi^2.
\end{equation}
Each $d_i$ contains the off-diagonal covariance and therefore represents a correlated share of the total change. A small negative radiative attribution can therefore occur even when the dipole rate barely changes. Expanding $\mathbf r=\mathbf r_{\rm SM}+\delta\mathbf O$ gives a linear term $2\mathbf r_{\rm SM}^T\Sigma^{-1}\delta\mathbf O$ and a nonnegative quadratic term. At $20$, $40$ and $80\,\mathrm{TeV}$ these are respectively $(11.5884,4.4168)$, $(3.2357,0.3443)$ and $(0.8936,0.0263)$. The linear dominance at large mass is driven mainly by a negative $\delta\epsilon_K$ on top of an SM central prediction below the data. The positive change quantifies the resulting deterioration relative to the SM reference.

The numerical supplement supplies the fixed input and covariance snapshot, the original reference matching source, two postprocessing scripts, expected JSON and CSV outputs, and a manifest of file hashes. The first script reconstructs the signed diagnostic and the explicitly labelled insertion, resonance and thermal proxies; the second reproduces Eqs.~\eqref{eq:mixingyardstick} and~\eqref{eq:radiativecomparison}. The postprocessing was rerun in a fresh extraction of the supplied package. Its numerical outputs agree with the supplied reference values to floating-point precision; final digits and runtime metadata can vary across environments. The check establishes the reproducibility of this postprocessing and the added comparisons. The original flavour-kernel source is supplied with the dependencies needed for a separate rerun; full dimuon matching remains a further calculation.

\end{document}